\documentstyle[preprint,eqsecnum,aps,prb]{revtex}

\begin{document}
\title{Epitaxial Growth of La$_{1/3}$Sr$_{2/3}$FeO$_3$ thin films by laser ablation}
\author{W. Prellier\thanks{%
prellier@ismra.fr} and B.\ Mercey}
\address{Laboratoire CRISMAT, CNRS\ UMR 6508,\\
6 Bd du Mar\'{e}chal Juin, 14050 Caen\\
Cedex, FRANCE.}
\date{\today}
\maketitle

\begin{abstract}
We report on the synthesis of high quality La$_{1/3}$Sr$_{2/3}$FeO$_3$
(LSFO) thin films using the pulsed laser deposition technique on both SrTiO$%
_3$ (STO) and LaAlO$_3$ (LAO) substrates (100)-oriented. From X-Ray
diffraction (XRD) studies, we find that the films have an out-of-plane
lattice parameter around 0.3865nm, almost independent of the substrate (i.e.
the nature of the strains). The transport properties reveal that, while LSFO
films deposited on STO exhibit an anomaly in the resistivity vs temperature
at $180K$ (corresponding to the charge-ordered transition and associated
with a transition from a paramagnetic to an antiferromagnetic state), the
films grown on LAO display a very small magnetoresistance behavior and
present an hysteresis around $270K$ under the application of a $4T$ magnetic
field. The changes in transport properties between both substrates are
discussed and compared with the corresponding single crystals.
\end{abstract}

\newpage

The transition metal oxides (TMO) have been extensively studied in the past
years, in particular due to the metal-insulator (MI) transition \cite{1}.
Among the different compounds, the manganites RE$_{1-x}$A$_x$MnO$_3$
(RE=rare earths, A=alkaline earths) have received a great interest these
last years due to their colossal magnetoresistance (MR) properties \cite{2,3}%
. An interesting feature of these materials is the charge-ordering phenomena 
\cite{4,5} i.e. the ordering of cations of different charges (Mn$^{3+}$/Mn$%
^{4+}$ for example) on specific lattice sites. This characteristic can be
seen in others compounds such as nickelates \cite{6} or layered manganites 
\cite{7}. In fact, it has also been recently observed in others perovskites
such as La$_{1/3}$Sr$_{2/3}$MnO$_3$ (LSFO) \cite{8}, which has been first
studied by Waugh {\it et al.} \cite{9}. Even if more studies appear very
recently on such materials \cite{10,11,11b,11c}, there are vey few reports
on thin films growth and to our knowledge, only for the particular
composition SrFeO$_{3-\delta }$ \cite{11b,11c}. Interestingly, theses
materials show properties close to the manganites and charge
disproportionation has also been investigated in the past \cite{12}.
Moreover, they are suitable candidates for MR properties \cite{13}. The
synthesis of charge-ordered (CO) manganite thin films is difficult owing to
the distortion of the lattice induce by the substrate \cite{14} and usually,
the properties of the CO thin films are different from the bulk. For
example, in the case of Pr$_{0.5}$Ca$_{0.5}$MnO$_3$, the CO state can not
been completely established. This means that the CO\ state is weaken due to
the substrate-induce strains because it is impossible to accomodate the
quite large changes of structural parameters when occuring below the CO
transition ($T_{CO}$) \cite{15}.

In an effort of understanding the particular substrate-effects of the CO
oxide compounds, it is interesting to look at others compositions. La$_{1/3}$%
Sr$_{2/3}$FeO$_3$ is such an example. For these reasons, we have synthesized
La$_{1/3}$Sr$_{2/3}$FeO$_3$ thin films on two types of substrates that
induce compression or tensile strains. We measured the structural lattice
parameters and transport properties of the films. Our results are report in
this communication.

Thin films of LSFO were grown using the pulsed laser deposition technique.
The target used has a nominal composition of La$_{1/3}$Sr$_{2/3}$FeO$_3$. In
the bulk form, LSFO has a rhombohedral structure ($R\overline{3}c$) with a
lattice constant $a_R=0.547nm$ and an edge angle $a_R=60.1{{}^{\circ }}$ at
room temperature \cite{11,12}. In fact, the average structure can be
considered as a cubic perovskite slightly distorted with a parameter close
to $a\approx 2a_P\approx 2\times 0.387nm=0.774nm$ (where $a_P$ being the
ideal parameter of the cubic perovskite). Thus, we expect to be able to grow
a thin film of this composition directly on a perovskite substrate. For
facilitating the reading of this paper, we keep the simple cubic cell
notation (index with C). Substrates used were (100)-SrTiO$_3$ (STO, cubic
with $a=0.3905nm$). Also, (100)-LaAlO$_3$ substrates (LAO, pseudocubic with $%
a=0.3788nm$) were utilized in order to compare the effect of tensile or
compressive stress. The substrates were kept at a constant temperature
around $710{{}^{\circ }}C$ during the deposition which was carried out at a
pressure of $200mTorr$ of flowing oxygen. After deposition, the samples were
slowly cooled to room temperature at a pressure of $500Torr$ of oxygen. The
deposition rate is $4Hz$ and the energy density is close to $2J/cm^2$.
Further details of the targets preparation and the deposition procedure are
given elsewhere \cite{15}. The structural study was done by X-Ray
diffraction (XRD) using a Seifert XRD 3000P for the $\Theta -2\Theta $ scans
and a X'Pert Phillips for the in-plane measurements (Cu, K$\alpha $ , $%
\lambda $=0.15406nm). The composition of the film was checked and
corresponds to the nominal one in the limit of the accuracy. Direct current
resistivity was measured by a four-probe method under magnetic fields up to
7T using a Quantum Design PPMS and magnetization was recorded using a
Quantum Design MPMS SQUID magnetometer as a function of the temperature in
the range 4-300K.

Fig.1a shows the $\Theta -2\Theta $ scan for an as-grown film deposited on
STO substrate. Two diffraction peaks can be seen corresponding to an
out-of-plane parameter of 0.3865nm. These two peaks can be indexed, based on
the cubic perovskite, as the (100)$_C$ and the (200)$_C$ reflections. Note
that the high intensity and the sharpness of the peaks, reveal a high
quality of the growth which is confirmed by the low value ($\approx 0.3{%
{}^{\circ }}$) of the full-width-at-half-maximum (FWHM) of the $\omega $%
-scan recorded around the (002)$_C$ (see inset of Fig.1a). Data used to
determine the in-plane lattice parameters of LSFO thin film were obtained
from asymmetric XRD scans of the STO (202) and (220) reflections. These
scans detected the (202)$_C$ and (220)$_C$ of LSFO. The unit cell is found
to be tetragonal with $a=0.3882nm$ and $c=0.3865nm$. These values are close
to the bulk value, referring to the cubic unit cell ($a\approx 0.387nm$) 
\cite{11,12}. In fact, the out--of-plane lattice parameter decreases
slightly and the in-plane lattice parameter increases lightly confirming
that the LSFO film is under tensile stress on STO\ substrate, keeping the
volume of the cell almost constant ($58.24\AA ^3$ in the film to be compared
with $57.96\AA ^3$ in the corresponding bulk material). Fig.1b shows a $\Phi 
$-scan obtained around the (202)$_C$ reflection. Four peaks separated from $%
90{{}^{\circ }}$, are visible and correspond to a four-fold symmetry. The
sharp peaks at $90{{}^{\circ }}$ intervals indicate a perfect in-plane
alignment with the substrate. Details investigations of the structure and
the microstructure are in progress using high resolution electron microscopy
and will be published elsewhere. The $\Theta -2\Theta $ scan for an as-grown
film deposited on LAO substrates is presented in Fig.2. As in the case of
STO, two diffraction peaks are clearly visible close to those of the
substrate, corresponding to an out-of-plane parameter of $0.386nm$.
Surprisingly, this value is very close to the one obtained on STO indicating
that the strains seem not to play an important role on the structure (or at
least on the lattice parameters) of these compounds. Even if lattice
parameters are similar for both substrates, the quality of the film\ are
different. This is evidenced by the full width at half-maximum (FWHM) of the 
$\omega $-scan recorded around the (002)$_C$ (see inset of Fig.2). On LAO,
the value is higher ($\approx 0.4{{}^{\circ }}$) on this substrate. The
rocking curves measurements characterize informations on the angular
distribution of the LSFO crystallites in the film. Thus, the narrow value on
STO as compared to LAO, shows that the $c$-axis parameter of LSFO films on
LAO has a spread mosaicity, i.e. there are not perfectly parallel to each
others.

We present in Fig.3 the temperature dependencies of the resistivity as well
as the magnetization for a LFSO film grown on STO. The resistivity at room
temperature is low ($\approx 10^{-4}\Omega .cm$) but it is still about one
order of magnitude higher than in the bulk \cite{11}. When the temperature
is decreasing, the resistivity is increasing almost gradually with a
semiconductorlike behavior. An anomaly around $180K$ is observed and
corresponds to the charge-ordered transition $T_{CO}$. At this temperature,
the magnetization shows a cusp which is characteristic of an
antiferromagnetic transition. However, at low temperature the magnetization
value is still very high but most probably a result of the substrate
component. Different features can be seen from these datas as compared with
the single crystals \cite{11}. First, the cusp in magnetization is weaker in
the film and second, there is no an abrupt change in resistivity as seen in
LSFO single crystals. This is not surprising because the charge-ordering
transition is never as pronounced in thin films \cite{14,15}, even if the
structure shows an ordering at low temperature \cite{15} and this is due to
the presence of substrate which does not allowed the change of lattice
parameters below the CO\ transition observed in the bulk. One can also
notice that $T_{CO}$ is $20K$ lower than the in the bulk ($200K$) . The
first explanation of the decreasing of the $T_{CO}$ is that, it is often
difficult to reach the optimized properties in thin films without a complex
growth process \cite{18}, mostly due to the substrate-induced strain which
bends the structure of the film. Another possibility might comes for the
oxygen stoechiometry since the volume of the cell of the film is increasing
of $0.3\AA ^3$. According to Dann {\it et al}., it would correspond to an
increase of the Sr content. Equivalently, this would correspond to an
increasing of the oxygen \cite{18b}. In our LSFO film, the changes of
lattice parameter confirm the tensile stress of the film and an increase of
the oxygen content led to a decrease of both lattice parameters (in-plane
and out-of-plane) \cite{18b}. Since in our case, only one parameter is
increasing, we think that the oxygen content is not the reason of such a
behavior.

Fig.4 shows the $\rho (T)$ without field and with a $7T$ magnetic field for
the LSFO film deposited on LAO. We can see that the film displays an
insulating behavior characteristic of the CO\ compound. However, in this
LSFO on LAO, we can notice that the film exhibit a small magnetoresistance
effect under a $7T$ magnetic field. Moreover, an anomaly around $260K$ with
a thermal hysteresis of few kelvins, are observed. These two features
correspond to a transition which is, in fact, close and higher to the $%
T_{CO} $ transition.

These results indicate that the CO state\ is weaker when deposited on\ STO\
rather than on LAO since the $T_{CO}$ is shifted to higher temperature in
the LAO case. In others words, the charge-ordered state becomes more stable
under compressive strain. Such feature as been observed recently in the case
of manganite thin films where the ferromagnetism is enhanced under tensile
stress \cite{15,19} (or equivalently the CO\ state becomes less stable under
tensile stress). The charge-ordered state seems to behave in the same way
with Mn and Fe based compounds. Further compositions need to be investigated
in the future to confirm theses results.

In conclusion, we have investigated the structural, transport and magnetic
properties of La$_{1/3}$Sr$_{2/3}$FeO$_3$ thin films grown by the pulsed
laser deposition technique. Highly crystallized films were obtained on both
substrates SrTiO$_3$ and LaAlO$_3$. We have seen that the out-of-plane
lattice parameter is almost independent of the type of strains. The films
show an anomaly in $\rho (T)$ and $M(T)$ close to the corresponding
charge-ordered transition observed in the bulk material. However, this
transition is lower in the case of STO\ and higher in the case of LAO, as
compared to the bulk compound confirming that a charge-oreder state in less
stable under tensile stress. We explained these differences by the stress
induced by the substrate upon the film.

\newpage

Figures Captions:

Figure 1a: Room temperature $\Theta -2\Theta $ XRD pattern of LSFO on STO.
Peaks labeled with the index S refer to the substrate peaks. The inset is
the rocking curve ($\omega $-scan) of (002)$_C$ reflection of the LSFO film.

Figure 1b: $\Phi $-scan of the (202)$_C$ family peaks in LSFO film showing a
very good in-plane orientation of the film.

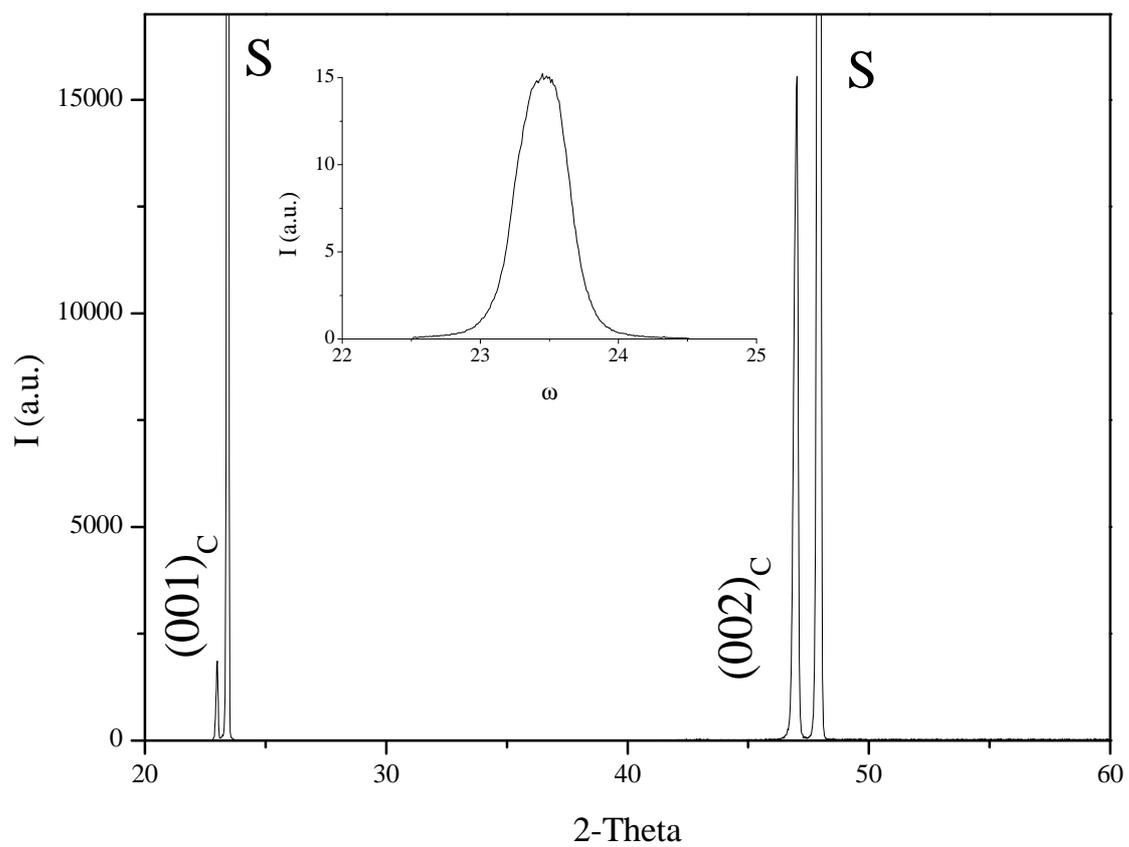
Figure 2: Room temperature $\Theta -2\Theta $ XRD pattern of LSFO on LAO.
Peaks labeled with the index S refer to the substrate peaks. The inset is
the rocking curve ($\omega $-scan) of (002)$_C$ reflection of the LSFO film.

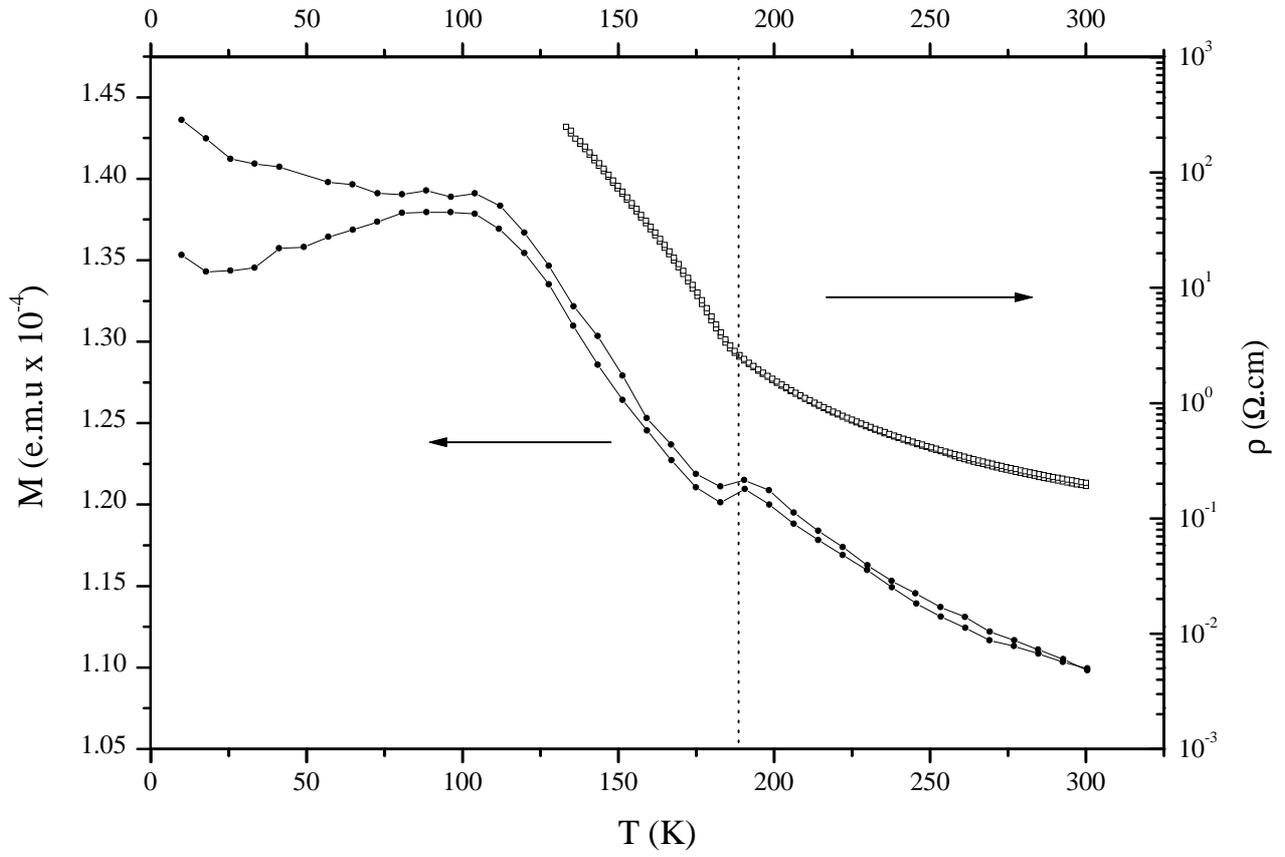
Figure 3: Temperature dependence of the resistivity $(\rho )$ and
magnetization $(M)$ at $\mu _0H=0.5T$ for a LSFO film on STO. No MR\ was
observed with a 7T\ magnetic field. The magnetic field is applied
perpendicular to the surface of the substrate.

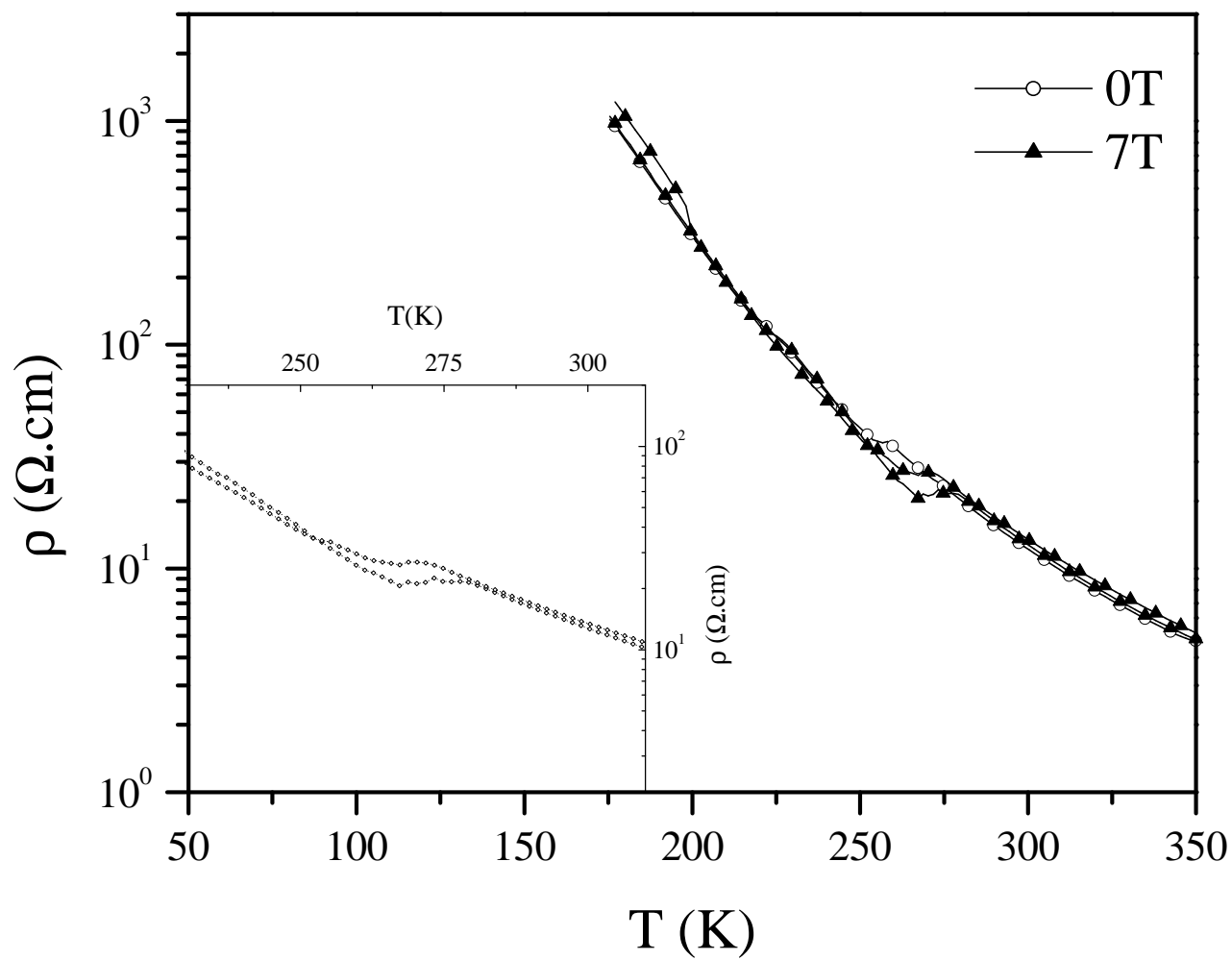
Figure 4: Temperature dependence of the resistivity $(\rho )$ for a LSFO on
LAO under different applied magnetic field. Note the small magnetoresistance
effect. The inset is magnified resistivity curve in the range $240-300K$
showing a clear phase transition around $270K$. The magnetic field is
applied perpendicular to the surface of the substrate.

\end{document}